\documentclass[aps,twocolumn,showpacs]{revtex4}
\usepackage{dcolumn}
\usepackage{graphicx}
\usepackage{amsmath}
\usepackage{amsfonts}
\usepackage{amssymb}
\usepackage{psfrag}
\usepackage{wrapfig}
\usepackage{subfigure}
\usepackage{makeidx}
\usepackage{bm}
\usepackage{epsf}
\usepackage{multirow}
\DeclareMathOperator{\sech}{sech}

\begin{document}

\title{Beating Effects of Vector Solitons in Bose-Einstein Condensate}
\author{Li-Chen Zhao}\email{zhaolichen3@nwu.edu.cn}
\address{$^{1}$School of Physics, Northwest University, Xi'an, 710069, China}
\address{$^{2}$Shaanxi Key Laboratory for Theoretical Physics Frontiers, Xi'an, 710069, China}

\date{\today}
\begin{abstract}

We study the beating effects of solitons in multi-component coupled Bose-Einstein condensate systems. Our analysis indicate that the period of beating behavior is determined by the energy eigenvalue difference in the effective quantum well induced by solitons, and the beating pattern is determined by the eigen-states of  quantum well  which are involved in the beating behavior. We show that the beating solitons correspond to linear superpositions of eigen-states in some quantum wells, and the correspondence relations are identical for solitons in both attractive interaction and repulsive interaction condensate. This provides a possible way to understand the beating effects of solitons for attractive and repulsive interaction cases in a unified way, based on the knowledge of quantum eigen-states. Moreover, our results demonstrate many different beating patterns for solitons in three-component coupled condensate, in sharp contrast to the beating dark soliton reported before. The beating behavior can be used to test the eigenvalue differences of some certain quantum wells, and   more abundant beating patterns are expected to exist in more components coupled systems.

\end{abstract}
\maketitle

\section{Introduction}

Multi-component Bose-Einstein condensate (BEC) provides a good platform to study dynamics of vector solitons \cite{Kevrekidis}. Many different vector solitons have been obtained in the two-component coupled BEC systems, such as bright-bright soliton \cite{Zhang}, bright-dark soliton \cite{Nistazakis}, dark-bright soliton \cite{XBO,Carr,BEC,PAG}, dark-dark soliton \cite{Ebd}. Bright-bright and bright-dark soliton usually exist in the coupled BEC with attractive interactions \cite{Lakshman,Lingdnls,Feng,DSWang}, while dark-bright and dark-dark soliton usually exist in the coupled BEC with repulsive interactions \cite{Hamner,Middelkamp,Yan}.
Recently, it was shown that it is possible to find dark-anti-dark soliton in BEC with unequal inter and intra-species interaction strengths \cite{Qu,Qu2,Danaila}.  All those previous reported solitons are stable and have no beating effects, but some of them can be used to generate beating solitons. For example, beating dark solitons were shown to exist in the two-component coupled BEC with equal inter and intra-species repulsive interactions \cite{park,Yanbd,Charalampidis}, which were generated from the dark-bright soliton with the $SU(2)$ symmetry property \cite{Kartashov}. Based on abundant vector solitons for more components cases, it is naturally expected that there should be more exotic beating patterns for more components coupled BEC systems.

On the other hand, there are also beating anti-dark solitons in two-component BEC with attractive interactions (see Fig. 1(a)), which can be seen from the results for rogue wave and breathers in coupled systems with attractive interactions \cite{Zhao,liu}. Then, we note that the beating patterns of dark soliton and anti-dark soliton  admit many similar properties, even though they exist in different interaction cases. Can we find a mechanism to understand the beating effects of them in a unified way? As far as we know, the beating effects have not been discussed systemically to uncover the underlying mechanisms. It is essential to discuss the beating patterns in details and find some fundamental mechanisms for these different beating behaviors.

In this paper, we study the beating effects of vector solitons in details. The analysis suggest that the beating effects of soliton is determined by the energy eigenvalue difference and corresponding eigen-states in induced quantum wells. The quantum wells admit identical forms for both attractive and repulsive interactions cases. In this way, we show that beating anti-dark solitons and beating dark soliton correspond to the same eigen-problems in a quantum well. Their beating period and pattern can be understood in a unified way, based on the well-known knowledge of linear superposition of quantum eigen-states. Furthermore, we demonstrate that there are some new beating patterns in three-component coupled BEC systems with the aid of $SU(2)$ and $SU(3)$ symmetry, such as beating bright soliton with double-hump, beating dark solitons with double-valley, in sharp contrast the well-known beating dark solitons. These results can be further extended to discuss more components involved cases, and the beating behaviors can be all understood well based on the knowledge of quantum eigen-states in quantum mechanics. The beating characters can be also used to test the energy eigenvalue differences of some certain quantum wells.

\begin{figure*}[htb]
\centering
\label{fig:1}
{\includegraphics[height=105mm,width=155mm]{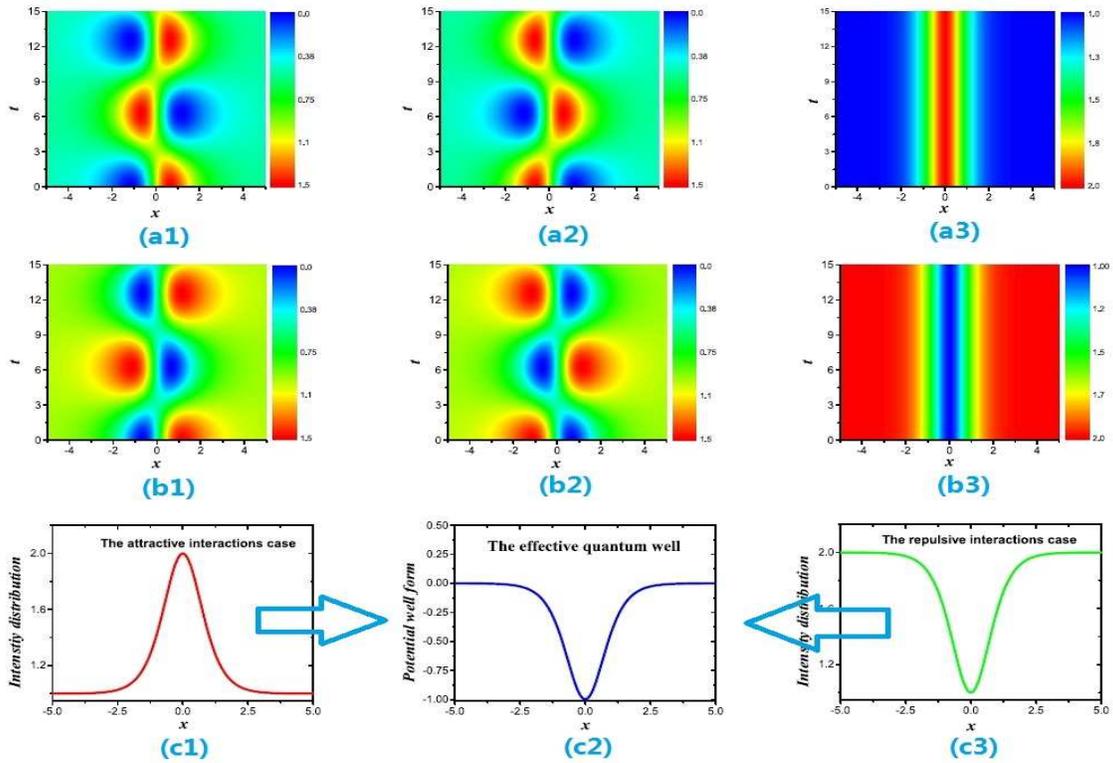}}
\caption{(color online)  (a1-a3) Beating anti-dark soliton for two-component coupled BEC with attractive interactions.  The three figures show the density evolution of component $\psi_1$, $\psi_2$ and the superposition of them  respectively. (b1-b3) Beating dark soliton for two-component coupled BEC with repulsive interactions. The three figures show the density evolution of component $\psi_1$, $\psi_2$ and the superposition of them  respectively. (c1-c3) show that the beating solitons in repulsive and attractive cases correspond to the superposition of  identical eigen-states in a quantum well $-f \ \sech^2[\sqrt{f} x]$. Our analysis indicates that the beating effects are induced by the coherence superposition of eigen-states in the quantum well. The beating period is determined by the energy eigenvalue difference in the quantum well. These characters holds well for the beating solitons in both attractive and repulsive interaction cases. The parameters in soliton solutions are $a=1$, and $f=1$. }
\end{figure*}

\section{The beating effects of two-component solitons}

We firstly study the simplest cases for vector solitons in a two-component coupled BEC. The dynamical equation can be written as the following dimensionless coupled model,  \begin{eqnarray}
&&i \frac{\partial \psi_1}{\partial t}+ \frac{1}{2} \frac{\partial^2 \psi_1}{\partial x^2} +\sigma (|\psi_1|^2+|\psi_2|^2) \psi_1=0,\nonumber\\
&&i \frac{\partial \psi_2}{\partial t}+ \frac{1}{2} \frac{\partial^2 \psi_2}{\partial x^2} + \sigma (|\psi_1|^2+|\psi_2|^2) \psi_2=0,
\end{eqnarray}
where  $\psi_1$ and $\psi_2$ denote the two component fields for the two-component coupled BEC systems \cite{Kevrekidis}. In this case, the inter and intra-species interactions have equal strength, and the model can be solved exactly by Darboux transformation, Hirota bilinear method, etc. Bright-bright and bright-dark soliton usually exist in the coupled BEC with attractive interactions $\sigma=1$ \cite{Lakshman,Lingdnls,Feng,DSWang}, while dark-bright and dark-dark soliton usually exist in the coupled BEC with repulsive interactions $\sigma=-1$ \cite{Hamner,Middelkamp,Yan}. The bright-bright soliton or dark-dark soliton can not be used to generated beating behavior, since the bright solitons or dark solitons in two components admit identical distribution profile and chemistry potential value. It has been shown that dark-bright and bright-dark soliton can be used to generate beating solitons. Dark-bright soliton can generate beating soliton, as shown in Fig. 1(b). It is seen that the beating behavior just emerge in each component, but there is no beating effects for the whole density of the two components, and the whole density profile is stable dark soliton. Therefore, it has been called beating dark soliton. The beating anti-dark soliton can be also generated from the well-known bright-dark solitons. Correspondingly, the whole density profile is an anti-dark soliton (shown in Fig. 1(a)), therefore it is called beating anti-dark soliton. To find a unified way to understand them, we represent them separately for the attractive and repulsive interactions cases. The beating anti-dark soliton can be given as  follows with $\sigma=1$,
\begin{eqnarray}
\psi_1&=&- (\sqrt{f+a^2} \  \sech[\sqrt{f} x] \ e^{i  f/2 t}+a\ \tanh[\sqrt{f} x])\nonumber\\
&&\frac{1}{\sqrt{2}}  e^{i a^2 t}, \\
\psi_2&=&- (\sqrt{f+a^2} \  \sech[\sqrt{f} x] \ e^{i  f/2 t}-a\ \tanh[\sqrt{f} x])\nonumber\\
&&\frac{1}{\sqrt{2}}  e^{i a^2 t}
\end{eqnarray}
where $a$ denotes the amplitude of plane wave background for dark soliton component. The beating dark soliton can be written as follows with $\sigma=-1$,
\begin{eqnarray}
\psi_1(x,t)&=&-( a \  \sech[\sqrt{f} x] \ e^{i  f/2 t}+\sqrt{f+a^2} \ \tanh[\sqrt{f} x])\nonumber\\
&&\frac{1}{\sqrt{2}}  e^{-i (a^2+f) t},\\
\psi_2(x,t)&=&-( a \  \sech[\sqrt{f} x] \ e^{i  f/2 t}-\sqrt{f+a^2} \ \tanh[\sqrt{f} x])\nonumber\\
&&\frac{1}{\sqrt{2}}  e^{-i (a^2+f) t},
\end{eqnarray}
where $\sqrt{f+a^2}$ denotes the amplitude of plane wave background for dark soliton component. The beating period is obviously determined by the chemical potential difference for solitons \cite{Charalampidis}.
The beating behaviors are distinctive for attractive and repulsive interactions cases. We would like to find a unified way to understand them, since  the beating behavior for dark soliton and anti-dark soliton always admit identical oscillation period if the parameter $f$ of solitons is chosen identically. Why this character holds for the two different cases, one for attractive case, and one fore repulsive case? Maybe there are many different ways to understand this point. We  would like to present one possible way to understand this, based on the relations between solitons and eigen-states in a quantum well. It will be shown that the classical linear superposition of eigen-state in quantum mechanics can explain perfectly the beating behavior of solitons in both attractive and repulsive cases.

 \begin{table}[!hbp]
 \centering
\begin{tabular}{|c|c|c|}
\hline
\ Solitons in AI  \ &\  Eigenvalues in a QW  \ & \ Solitons in RI  \ \\
\hline
dark soliton & $0$ & dark soliton \\
  bright soliton & $-f/2$ & bright soliton \\
\hline
\end{tabular}
\caption{The correspondence between soliton states in two-component BEC and eigen-states in a quantum well $-f \ \sech^2[\sqrt{f} x]$. It is seen that the solitons in attractive and repulsive cases correspond to identical eigenvalues in the quantum well. The beating period is surely determined by the eigenvalue difference, which can be understood well by the knowledge in quantum mechanics. ``AI", ``QW", and ``RI" denote attractive interaction BEC, quantum well, and repulsive interaction BEC respectively. }
\end{table}

Calculating that $|\psi_1|^2+|\psi_2|^2=a^2+f \ \sech^2[\sqrt{f} x]$   in two-component coupled NLS with attractive case,  and then substituting this to the equation (1) and (2) with $\sigma=1$, one can find that the beating soliton solution is related with eigen-problem in a quantum well $-f \sech^2[\sqrt{f} x]$. The corresponding eigen-equation is
\begin{eqnarray}
&& - \frac{1}{2} \frac{\partial^2 \psi_j}{\partial x^2} - f \sech^2[\sqrt{f} x] \psi_j=\mu_j \psi_j .
 \end{eqnarray}
 We can prove directly that the bright soliton and dark soliton, which are used to generate beating anti-dark soliton,  are both the eigen-states in the quantum well $-f \sech^2[\sqrt{f} x]$.  It is seen that the bright soliton corresponds to the eigenvalue $-f/2$ in the quantum well, and the dark soliton corresponds to eigenvalue zero in the quantum well. This agrees well with the results in quantum wells \cite{Landau,Rosen}. For the beating dark soliton in the  repulsive case, $|\psi_1|^2+|\psi_2|^2= a^2+ f\   \tanh^2 [\sqrt{f} x] $  can be used to find related eigen-states in a quantum well for the repulsive cases. With the help of $tanh^2(x)=1- \sech^2(x)$, we can rewrite the potential form as  $a^2+ f\   \tanh^2 [\sqrt{f} x] =a^2+f\ -f\ \sech^2[\sqrt{f} x] $.  After transferring the constant terms to chemistry potential terms, the effective quantum well also becomes $-f \sech^2[\sqrt{f} x]$. Interestingly, \emph{the eigen-problem for repulsive case is identical with the one in attractive case, namely, they admit the identical eigen-equation}. This is demonstrated in Fig. 1(c). It is seen that the eigenvalues of bright soliton and dark soliton, which are used to generate beating dark soliton, are also $-f/2$ and $0$ in the quantum well. These characters are summarized in Table I. In this way, we establish the correspondence between solitons  and eigen-states in quantum well. This provides possibilities to explain beating effects of soliton based on the knowledge of eigen-states in quantum mechanics.

\begin{figure*}[htb]
\centering
\label{fig:1}
{\includegraphics[height=105mm,width=155mm]{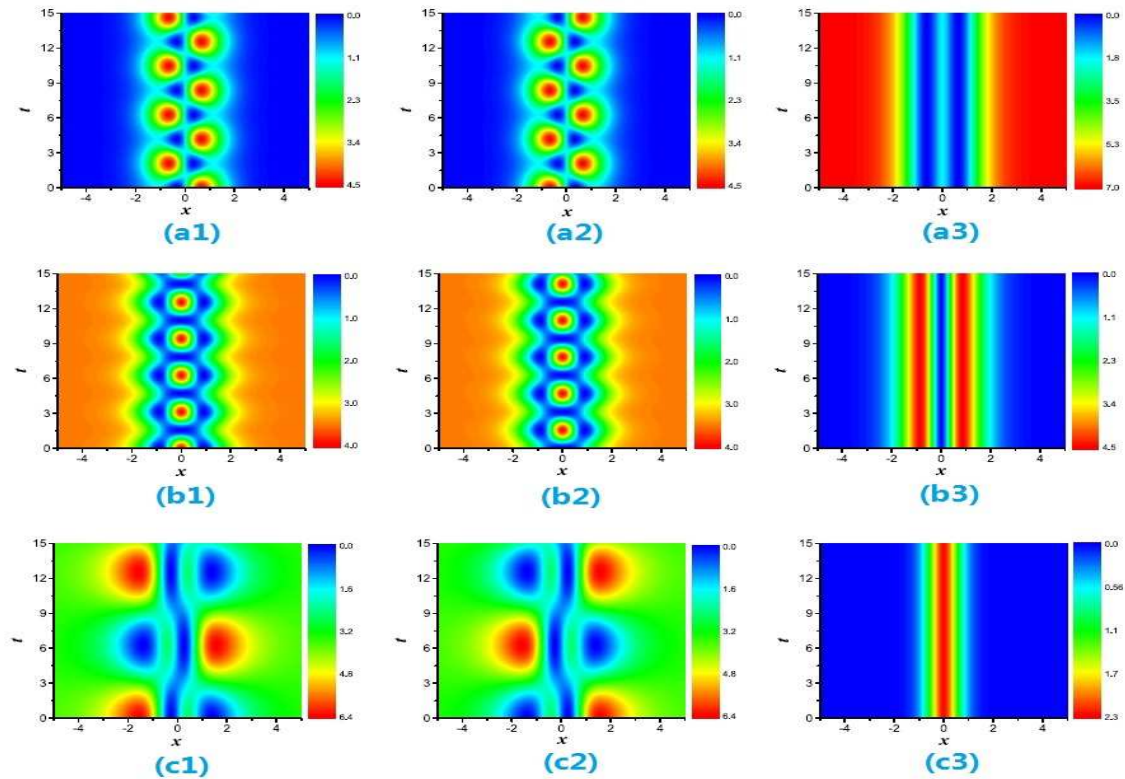}}
\caption{(color online) The beating solitons in three-component coupled BEC with repulsive interactions, which are generated from $SU(2)$ symmetry. (a1-a3) show the evolution of beating bright soliton with double-hump in the three components  respectively, which are superpositions of the ground state and the first-excited states in the quantum well $-3 f \ \sech^2[\sqrt{f} x]$.  (b1-b3) show the evolution of beating dark soliton with double-valley in the three components  respectively, which are superpositions of the ground state and the second-excited states in the quantum well. (c1-c3) show the evolution of beating dark soliton in the three components  respectively, which are superpositions of  the first-excited states and the second-excited states in the quantum well. The parameters in soliton solutions are $a=2$, and $f=1$. }
\end{figure*}

Therefore, the beating solitons are fundamentally a linear superposition form of eigen-states in a quantum well. In  quantum mechanics, arbitrary linear superposition of eigen-states are always the solution of the linear  Sch\"{o}dinger equation. But there is also bit difference that the linear superposition coefficients can not be arbitrary for Eq. (1) and (2), since they are nonlinear partial equation. The linear superposition of two eigen-states surely will admit beating behavior, and the beating period is determined by the eigenvalue difference \cite{Landau}. Therefore, \emph{there is no beating effects for superpositions of degenerated solitons, for which the solitons admit identical eigenvalue. }This provides a good way to understand the beating effects of solitons in BEC.

Most previous experiments in BEC demonstrate two-component solitons can be produced well by different density and phase modulation techniques \cite{BEC,Hamner,Middelkamp,Yan}.  Very recently, three-component soltion states were observed in a spinor BEC system \cite{Bersano}. Motivated by these developments, we would like to discuss the beating effects of solitons in three-component coupled BEC.

\section{The beating effects of three-component solitons}
The three-component coupled BEC can be described by the following dynamical equations,
 \begin{eqnarray}
&&i \frac{\partial \psi_1}{\partial t}+ \frac{1}{2} \frac{\partial^2 \psi_1}{\partial x^2} +\sigma (|\psi_1|^2+|\psi_2|^2+|\psi_3|^2) \psi_1=0,\nonumber\\
&&i \frac{\partial \psi_2}{\partial t}+ \frac{1}{2} \frac{\partial^2 \psi_2}{\partial x^2} + \sigma (|\psi_1|^2+|\psi_2|^2+|\psi_3|^2) \psi_2=0,\nonumber\\
&&i \frac{\partial \psi_3}{\partial t}+ \frac{1}{2} \frac{\partial^2 \psi_3}{\partial x^2} + \sigma (|\psi_1|^2+|\psi_2|^2+|\psi_3|^2) \psi_3=0,
\end{eqnarray}
where  $\psi_1$, $\psi_2$ and $\psi_3$ denote the three component fields for the three-component coupled BEC systems \cite{Kevrekidis}.
Most of previous reported three-component solitons are degenerated soliton or partly degenerated soliton  \cite{Lakshman,Lingdnls,Feng,Bersano}, such as bright-bright-bright soliton, bright-dark-dark soliton, dark-bright-bright soliton, and dark-dark-dark soliton, etc. For example, for dark-bright-bright soliton,  the dark soliton component admits one node and the two bright soltion components admit identical eigen-state (they are degenerated). The superposition of degenerated soliton can not produce any beating effects, and the non-degenerated soliton can produce beating effects which is similar to the ones in two-component case \cite{Lakshman,Lingdnls,Feng,Bersano}. This is because the dark soliton and bright soliton in partly degenerated solitons for three-component case, which correspond to identical eigen-states in the quantum well with the ones for two-component case. Therefore, we do not show them in details, and these characters can be seen in previously reported soliton solutions.

In fact, the vector soliton in the three-component coupled systems can admit non-degenerated solitons \cite{NA,NA2}. The dark soliton in one component can admit double-valley structure, and one bright soliton component can also admit a node which makes the bright soliton admits double-hump. These are quite different from the ones observed in \cite{Bersano}. Similarly, we can generated beating soliton from the eigen-states in a quantum well $-3 f  \ \sech^2[\sqrt{f} x]$. It is found that the three-component soliton correspond to identical eigen-states in the quantum well for attractive and repulsive interactions cases. Namely, the eigen-states of the linear  Sch\"{o}dinger equation
\begin{eqnarray}
&&- \frac{1}{2} \frac{\partial^2 \psi_j}{\partial x^2} - 3f  \sech^2[\sqrt{f} x] \psi_j=\mu_j \psi_j,
 \end{eqnarray}
can be used to generate three-component solitons for both attractive and repulsive interactions cases. This makes the beating patterns for attractive interactions case are similar to the ones in repulsive case. Therefore, we mainly discuss the beating solitons in three-component coupled BEC with repulsive interactions. Similar behaviors are expected in attractive case.

\begin{figure*}[htb]
\centering
\label{fig:1}
{\includegraphics[height=45mm,width=155mm]{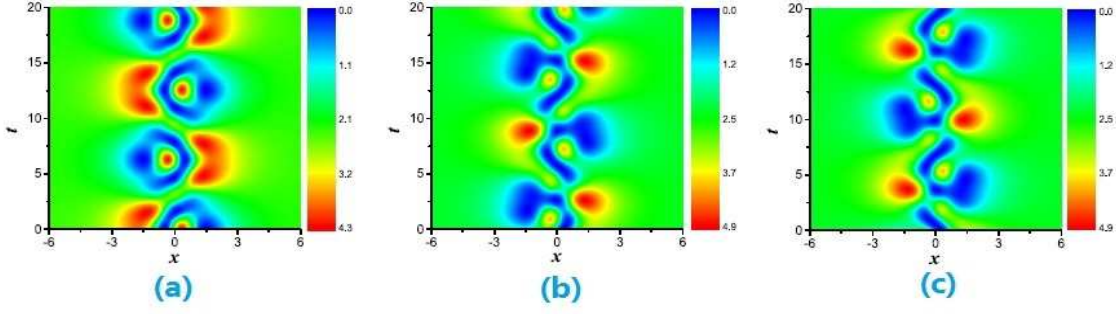}}
\caption{(color online) The beating dark soliton in three-component coupled BEC with repulsive interactions, which are generated from $SU(3)$ symmetry. (a-c) show the evolution of beating dark solitons in the three components respectively. The beating effects come from the superposition of ground state, the first-excited state and the second-excited state in the quantum well $-3 f \ \sech^2[\sqrt{f} x]$.  The parameters in soliton solutions are $a=2$, and $f=1$. }
\end{figure*}

The eigen-states $ \sech^2[\sqrt{f} x] $ ,  $ \sech[\sqrt{f} x]\ \tanh[\sqrt{f} x] $, and $(1-3 \tanh ^2[\sqrt{f} x]  $ correspond to the eigenvalue $-2f$, $-f/2$, and $0$  in the quantum well $-3 f  \ \sech^2[\sqrt{f} x]$ respectively. From the nodes of the eigen-states, we can know that the eigen-states, corresponding to eigenvalue $-2f$, $-f/2$, and $0$, are the ground state, the first-excited state, and the second-excited state respectively. This can be used to construct non-degenerated vector solitons for three-component coupled nonlinear Sch\"{o}dinger equation with attractive or repulsive interactions. Since the nonlinear equations have some additional constrain conditions on the coefficients of three wave functions,  we introduce some new coefficients for them, namely, $\phi_1(x)= a_3 \  \sech^2[\sqrt{f} x] $, $\phi_2(x)= b_3 \  \sech[\sqrt{f} x] \ \tanh[\sqrt{f} x] $, and  $\phi_3(x)=c_3(1-3 \tanh ^2[\sqrt{f} x]  $ .  We can identify the values of $a_3$ , $b_3$ and $c_3$ with the constrain condition $|\psi_1|^2+|\psi_2|^2+|\phi_3(x)|^2=a^2+3 f\   \tanh^2 [\sqrt{f} x] $.  Then the static vector non-degenerated  soliton of the three-component BEC with repulsive interactions $\sigma=-1$ can be given as, $\psi_{1s}=\frac{1}{2}\sqrt{ 3 (a^2-f)} \  \sech^2[\sqrt{f} x]\ e^{-i t  (a^2+f )}$, $\psi_{2s}=\sqrt{3  (a^2+2 f )}  \tanh [\sqrt{f} x]\   \sech[\sqrt{f} x] e^{-i t  (a^2+\frac{5 f}{2} )}$, and
$\psi_{3s}=\frac{1}{2}\sqrt{  (a^2+3 f)}\ (1-3 \tanh ^2[\sqrt{f} x] )\ e^{-i t  (a^2+3 f )}$. From the nodes of these eigen-states, we can know that $\psi_{1s}$, $\psi_{2s}$, and  $\psi_{3s}$ are the ground state, the first-excited state, and the second-excited state respectively.
The linear superposition of them can generate many different beating solitons, with the aid of  the $SU(2)$ or $SU(3)$ symmetry admitted by the three-component coupled nonlinear equations.

For $SU(2)$ symmetry case, the transformation matrix have many different forms. The beating patterns of them are similar, whose beating periods are identical.  As an example, we choose   $S_{3\times 3}=\left(
\begin{array}{ccc}
 -\sqrt{\frac{1}{2}} & -\sqrt{\frac{1}{2}} & 0  \\
 -\sqrt{\frac{1}{2}} & \sqrt{\frac{1}{2}} & 0 \\
 0 & 0 & 1 \\
\end{array}
\right)$, the linear transformation $S_{3\times3} \left(\psi_{1s},\psi_{2s},\psi_{3s}\right)^T$ (``T" denotes the transpose of a matrix) can be used to construct many different beating solitons.  Firstly, $S_{3\times3} \left(\psi_{1s},\psi_{2s},\psi_{3s}\right)^T$ describes a superposition of ground state and the first-excited state in the quantum well. The dynamical process for beating soliton are shown in Fig. 2 (a). It is seen that the beating behaviors are demonstrated on zero background, in contrast the the beating dark soliton and beating anti-dark soliton shown above. The superposition of them is a bright soliton with double-hump, therefore this beating soliton is call as beating bright soliton with double-hump. This is similar to the case which the beating dark soliton is named. Secondly, $S_{3\times3} \left(\psi_{1s},\psi_{3s},\psi_{2s}\right)^T$ describes a superposition of ground state and the second-excited state in the quantum well. The dynamical process for beating soliton are shown in Fig. 2 (b). It is seen that the beating behaviors emerge on a plane wave background, which is similar to the beating dark soliton discussed before. But there is a sharp difference, namely, the superposition of the beating components is a dark soliton with double-valley. Therefore, this beating soliton is called as beating dark soliton with double valley, in contrast to the beating dark soliton. Thirdly, $S_{3\times3} \left(\psi_{2s},\psi_{3s},\psi_{1s}\right)^T$ describes a superposition of the first-excited  state and the second-excited state in the quantum well. The dynamical process for beating soliton are shown in Fig.2 (c). It is seen that beating patterns also emerge on plane wave background, and the superposition of them is dark soliton. The dark soliton also admits one valley, therefore this beating soliton is a beating dark soliton. It should be noted that the beating pattern is different from the beating dark soliton in two-component case, since the eigen-states are different from the ones in two-component case.

For $SU(3)$ symmetry, the transformation matrix can be chosen   $S_{3\times 3}=\left(
\begin{array}{ccc}
 \sqrt{\frac{1}{3}} & \sqrt{\frac{1}{3}} & \sqrt{\frac{1}{3}} \\
 \sqrt{\frac{1}{3}} & -\sqrt{\frac{1}{3}} \exp \left(\frac{i \pi }{3}\right) & \sqrt{\frac{1}{3}} \exp \left(\frac{i 2 \pi }{3}\right)  \\
 \sqrt{\frac{1}{3}} & \sqrt{\frac{1}{3}} \exp \left(\frac{i 2 \pi }{3}\right) & -\sqrt{\frac{1}{3}} \exp \left(\frac{i \pi }{3}\right)   \\
\end{array}
\right)$ as an example without losing generality, the linear transformation $S_{3\times3} \left(\psi_{1s},\psi_{2s},\psi_{3s}\right)^T$ describes the superposition of ground  state, the first-excited state, and the second-excited state. The dynamical process for beating soliton are shown in Fig. 3.  The beating pattern become more complicated, since the beating period involves more periodic functions in this case. The superposition of them is  a dark soliton which admit one valley. Therefore, it is also a beating dark soliton, but its beating pattern is different from all previously reported ones. This suggests that more beating patterns can be found in more components involves cases, since more components coupled BEC correspond to deeper quantum wells, which admit more eigen-states with many different nodes.

\section{Conclusion and discussion}
In summary, we show that the beating patterns of soliton are determined by the eigenvalue difference and corresponding eigen-states in some effective quantum wells.  Especially, the effective quantum wells admit identical forms for both attractive and repulsive interactions cases. In this way, we show that beating anti-dark soliton and beating dark soliton correspond to the same eigen-problems in a quantum well, for two-component coupled BEC. Their beating period and pattern can be understood in a unified way, based on the well-known knowledge of linear superposition of quantum eigen-states. These characters holds well for more components involved cases. For an example, we demonstrate that there are some new beating patterns in three-component coupled BEC systems, such as beating bright soliton with double-hump, beating dark solitons with more humps or valleys, in sharp contrast the beating dark solitons reported before. The discussions on beating solitons can be extended to more components coupled BEC systems. It is expected that more components coupled BEC admit more different eigen-states in the effective quantum wells, which would induce more different beating patterns for vector solitons.

From the results of beating effects, we can know that the beating period is determined by the corresponding energy eigenvalue difference in the effective quantum wells. On the other hand, one can produce the initial density distribution and phase for beating solitons in BEC systems through the well-developed  density and phase modulation techniques \cite{BEC,Hamner,Middelkamp,Yan,Bersano}. The beating period would be measured directly in real experiments \cite{Carr2}. Then the beating period $T=\frac{2\pi}{\Delta}$, where $\Delta$ denotes the energy eigenvalue difference in some certain quantum wells. Therefore,  it is possible to measure the energy eigenvalue difference in many different quantum wells in multi-component BEC systems.

\section*{Acknowledgments}
This work is supported by National Natural Science Foundation of
China (Contact No. 11775176), Shaanxi Association for Science and Technology (Contact No. 20160216), The Key Innovative Research Team of Quantum Many-Body Theory and Quantum Control in Shaanxi Province (Grant No. 2017KCT-12), and the Major Basic Research Program of Natural Science of Shaanxi Province (Grant No. 2017ZDJC-32).

\end{document}